\documentclass[apj]{emulateapj}

\usepackage{times}

\def\ion#1#2{#1\,{\sc\romannumeral #2}}

\shorttitle{Observations of Reconnecting Flare Loops}
\shortauthors{Warren, O'Brien, \& Sheeley}

\begin{document}


\title{Observations of Reconnecting Flare Loops with the Atmospheric Imaging Assembly (AIA)}

\author{Harry P. Warren, Casey M. O'Brien\altaffilmark{1}, Neil R. Sheeley, Jr.}

\affil{Space Science Division, Naval Research Laboratory, Washington, DC 20375}

\altaffiltext{1}{Massachusetts Institute of Technology, Cambridge, MA 02139 USA.}


 \begin{abstract}
   Perhaps the most compelling evidence for the role of magnetic reconnection in solar flares
   comes from the supra-arcade downflows that have been observed above many post-flare loop
   arcades.  These downflows are thought to be related to highly non-potential field lines that
   have reconnected and are propagating away from the current sheet. We present new
   observations of supra-arcade downflows taken with the Atmospheric Imagining Assembly (AIA)
   on the \textit{Solar Dynamics Observatory} (\textit{SDO}). The morphology and dynamics of
   the downflows observed with AIA provide new evidence for the role of magnetic reconnection
   in solar flares. With these new observations we are able to measure downflows originating at
   much larger heights than in previous studies. We find, however, that the initial velocities
   measured here ($\sim$144\,km~s$^{-1}$) are well below the Alfv{\'e}n speed expected in the
   lower corona, and consistent with previous results. We also find no evidence that the
   downflows brighten with time, as would be expected from chromospheric evaporation. These
   observations suggest that simple two-dimensional models cannot explain the detailed
   observations of solar flares.
 \end{abstract}

\keywords{Sun: corona}


 \section{Introduction}

 Magnetic reconnection is a fundamental process in astrophysical plasmas and is believed to be
 responsible for a wide range of solar phenomena, including solar flares \cite[for a review
 see][]{zweibel2009}. Unfortunately, magnetic reconnection occurs on spatial scales that are
 too small to be resolved with current solar instrumentation and cannot be observed
 directly. There is, however, considerable indirect evidence that magnetic reconnection plays a
 central role in solar flares. Perhaps the most compelling piece of indirect evidence is the
 supra-arcade downflows that have been observed above many post-flare loop arcades
 \citep[e.g.,][]{savage2011,savage2010,sheeley2004,asai2004,innes2003,mckenzie1999}. These
 downflows appear to be related to highly non-potential field lines that have reconnected and
 are propagating away from the current sheet.

 Supra-arcade downflows have been observed with a number of instruments and their properties
 are summarized in the comprehensive review by \cite{savage2011}. The downlows often appear as
 dark, collapsing features that look like small ``tadpoles'' (voids with a trailing dark
 tail). Some downflows have a loop-like appearance while others appear to evolve from voids
 into loops. The downflows generally begin at relatively high velocities (100--200\,km~s$^{-1}$
 is typical) and then quickly decelerate to about 4\,km~s$^{-1}$. Observations at high spatial
 resolution give the impression that much of high temperature plasma in a flare is formed from
 collapsing loops \citep{sheeley2004}, even when individual features cannot be tracked. Similar
 features are observed in the outer corona when the streamer belt is viewed face on
 \cite[e.g.,][]{wang1999,sheeley2001a,sheeley2001b}.

 Despite the extensive analysis of previous observations a number of questions remain regarding
 the downflows and their relationship to magnetic reconnection. The initial velocities measured
 for the downflows tend to be much smaller than the estimated Alfv{\'e}n speed in the lower
 corona ($\sim$1000\,km~s$^{-1}$), the velocity expected for reconnection outflows. It is
 possible that the downflows are formed at large heights and are observed only after they have
 experienced some deceleration. In most models of solar flares the release of energy in the
 corona ultimately drives chromospheric evaporation and leads to the formation of hot, dense
 loops \cite[e.g.,][]{fisher1985,mariska1989}. Many of the downflows, however, are observed as
 dark features and it is not clear that all of the high temperature emission in the flare is
 formed from descending loops.  Finally, why some downflows appear as voids and while others
 appear as loops is unclear. It is possible that the observed morphology is strongly influenced
 by the viewing angle.

 The launch of the Atmospheric Imaging Assembly (AIA) on the \textit{Solar Dynamics
   Observatory} (\textit{SDO}) provides a new opportunity to investigate the dynamics of newly
 reconnected flux tubes in solar flares. AIA has several advantages over previous solar EUV
 imagers, including a higher cadence (12\,s) and more channels that observe at flare
 temperatures. The higher cadence allows us to create data sets optimized for observing the
 downflows at large heights. The AIA 94\,\AA\ and 131\,\AA\ channels observe emission from
 \ion{Fe}{18} and \ion{Fe}{21}, respectively. These lines are formed close to 10\,MK, the
 temperature at which the flare emission measure generally peaks. 

 Observations from the twin \textit{STEREO} spacecraft \citep{howard2008} can also be applied
 to this problem. During the 2010--2012 time frame the spacecraft are approximately 90$^\circ$
 from the Earth-Sun line, which allows flares to be observed simultaneously as limb events by
 AIA and disk events by the EUVI imagers on \textit{STEREO}.

 In this paper we report new observations of supra-arcade downflows taken with AIA and
 EUVI. With the improved capabilities of AIA we are able to track features at larger heights
 and in much weaker events than in previous studies. The combined AIA and EUVI observations
 provide compelling evidence that the downflowing voids and loops are manifestations of the
 same phenomenon observed along different lines of sight. We find, however, that the initial
 velocities are consistent with previous measurements and well below the estimated Alfv{\'e}n
 speed.  The relationship between the brightest emission in the flare and the downflows is also
 not resolved. We find no evidence that the emission from the downflows rises with time, as
 would be expected from chromospheric evaporation. Also, for these events the initial high
 temperature flare loops are observed to form almost ``in place'' with very little associated
 downward motion. These difficulties suggest that simple two-dimensional models of magnetic
 reconnection cannot be applied directly to detailed observations of solar flares.

 
 \section{Observations}

 \begin{figure}[h!]
 \centerline{\includegraphics[clip,scale=0.5]{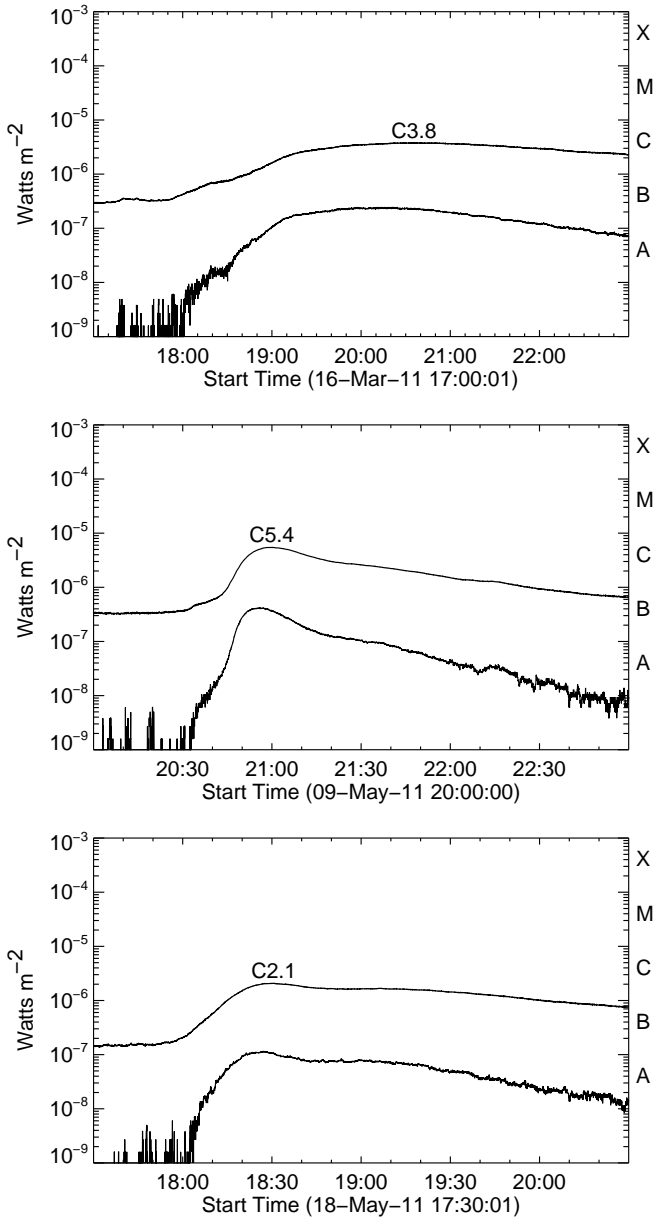}}
 \caption{Soft X-ray light curves observed with \textit{GOES} 15 for three limb
   flares. The May 9 and May 18 events are partially occulted.}
 \label{fig:goes}
 \end{figure}

 The AIA instrument on \textit{SDO} is a multilayer imaging telescope.  A full description of
 the instrument is given in \cite{lemen2011}. In brief, the AIA instrument images the full Sun
 in 10 different channels with 0.6\arcsec\ pixels at a typical cadence of 12\,s. There are 7
 EUV channels for imaging emission from the transition region and corona. The AIA 94\,\AA,
 131\,\AA, and 193\,\AA\ channels observe flare emission from \ion{Fe}{18}, \ion{Fe}{21}, and
 \ion{Fe}{24}, respectively. Each of the channels is also sensitive to emission formed at
 cooler temperatures. The 193\,\AA\ channel includes \ion{Fe}{12} 195\,\AA, which is an
 intrinsically bright line and makes \ion{Fe}{24} difficult to observe in smaller events. The
 lower temperature response for the 131\,\AA\ channel is primarily from \ion{Fe}{8} emission
 lines that are much weaker than \ion{Fe}{12} 195\,\AA, making the high temperature emission
 much easier to observe. The lower temperature lines in the 94\,\AA\ bandpass are largely
 unidentified, but they are also relatively weak. For a more complete description of the AIA
 temperature response see \cite{odwyer2010}.

 All of the AIA data that we have analyzed are sub-fields of the full disk images obtained from
 the LMSAL cutout service. The data has been processed using \verb+aia_prep+ to remove any
 energetic particle spikes and co-register the images from the different wavelengths to a
 common plate scale.

 For this work we consider three flares that occurred at the solar limb as viewed from Earth: a
 C3.8 event that began at about 18:00 UT on 2011 March 16, a C5.4 flare that began at about
 20:30 UT on 2011 May 9, and a C2.1 flare that begin at about 18:00 UT on 2011 May 18. The soft
 X-ray light curves from the \textit{GOES} 15 spacecraft are shown in
 Figure~\ref{fig:goes}. The \textit{GOES} light curves indicate that these flares are all long
 duration events that evolve over many hours.

\begin{figure*}[th!]
 \centerline{\includegraphics[clip,scale=0.45]{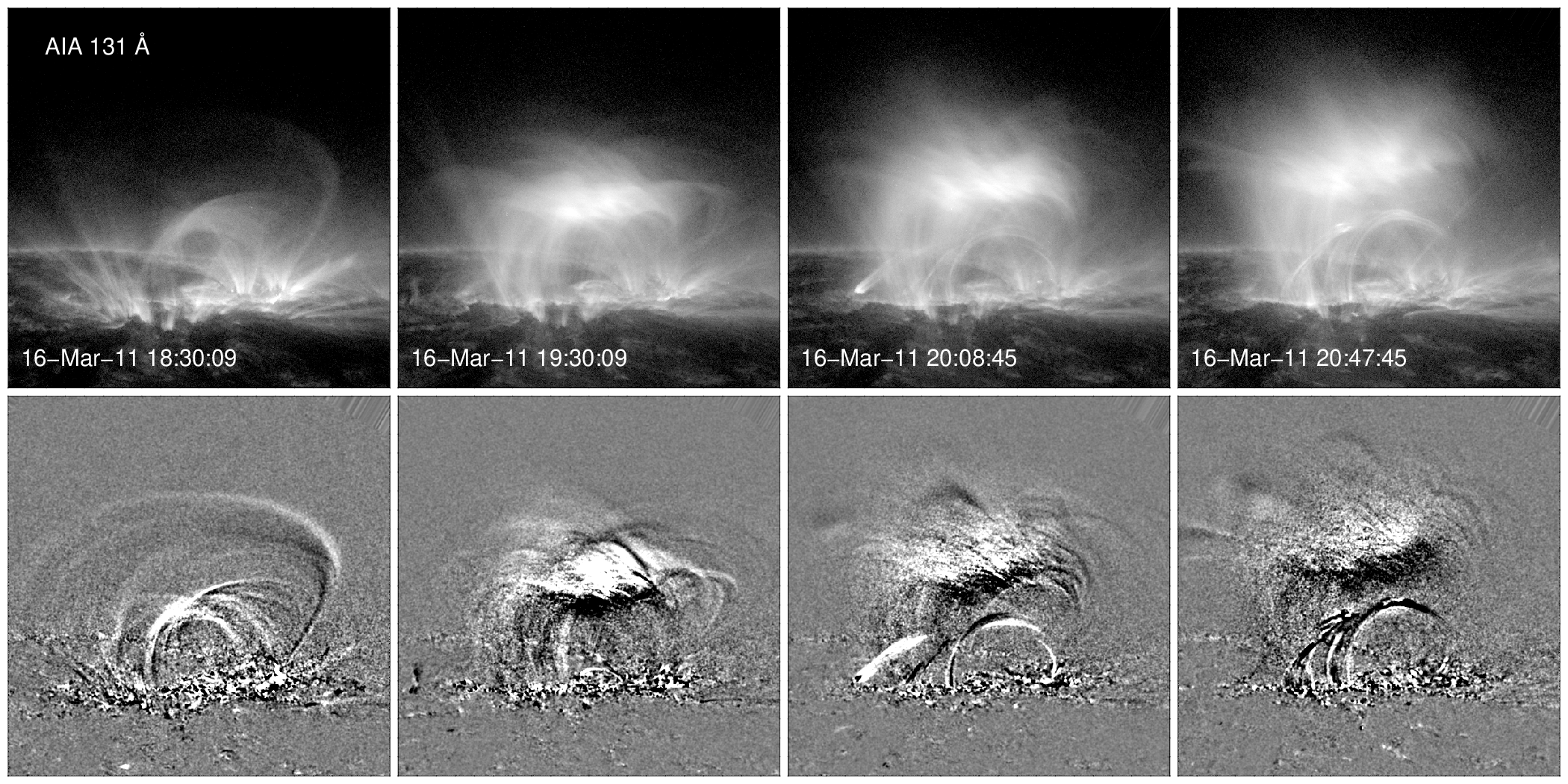}}
 \centerline{\includegraphics[clip,scale=0.45]{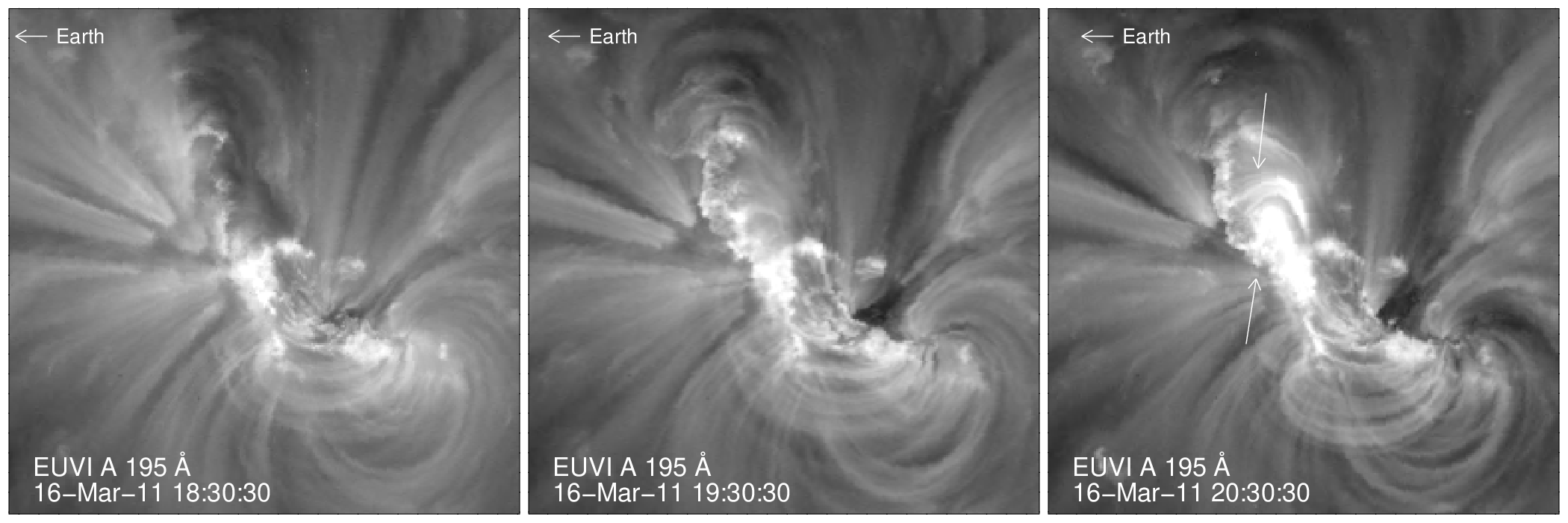}}
 \caption{Observations of a limb flare with AIA and EUVI on 2011 March 16. The top panels show
   the AIA 131\,\AA\ intensities scaled logarithmically at various times during the event. The
   middle panels show the running difference images for these times. The field of view of the
   AIA images is $350\arcsec\times319\arcsec$ and the image has been rotated. Details on how
   the running difference images are computed are given in the text. The bottom panels show the
   EUVI images from the A spacecraft. The size of the EUVI field of view is
   $408\arcsec\times408\arcsec$. The arrows in the final panel of the EUVI images indicates the
   approximate orientation of the flare ribbons. The electronic version of the manuscript
   contains a movie of the running difference images.}
 \label{fig:stereo0}
 \end{figure*}

\begin{figure*}[th!]
 \centerline{\includegraphics[clip,scale=0.45]{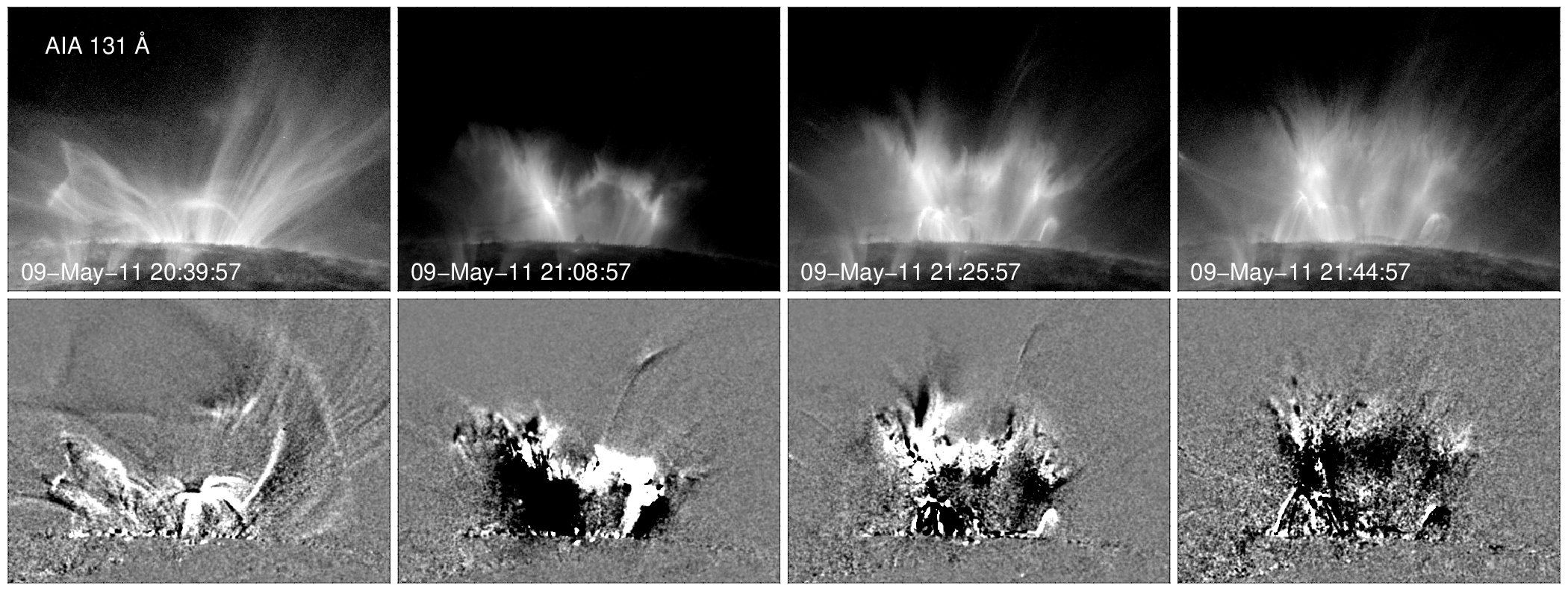}}
 \centerline{\includegraphics[clip,scale=0.45]{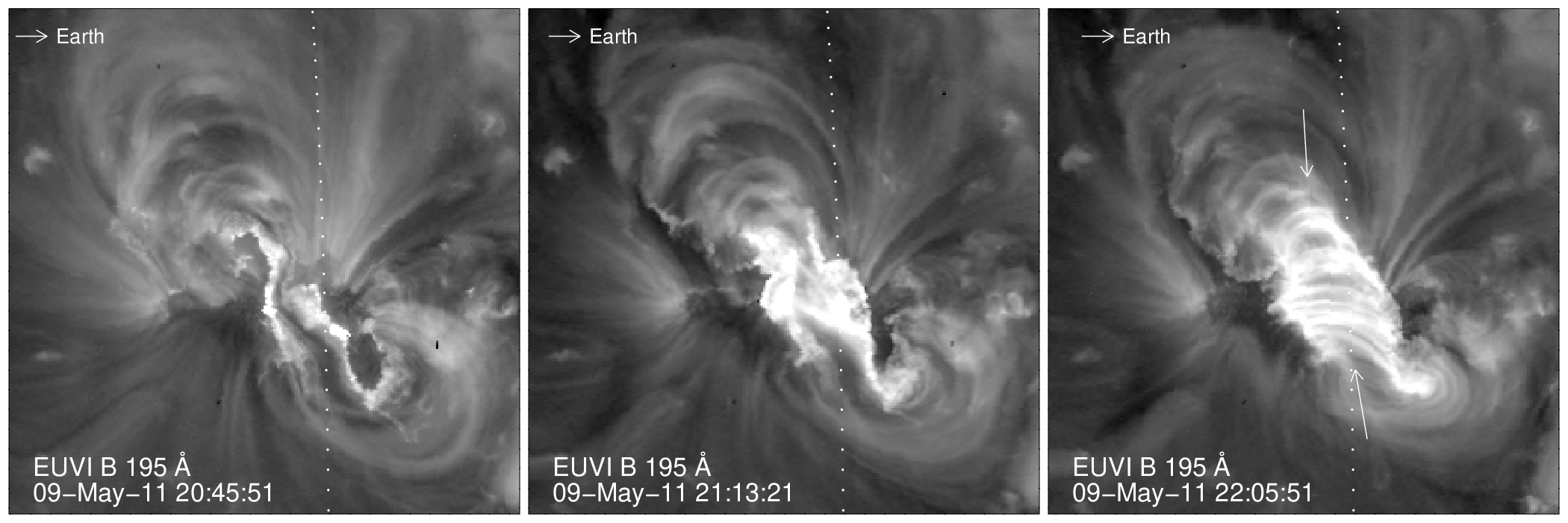}}
 \caption{The same as in Figure \protect{\ref{fig:stereo0}}, but for the 2011 May 9 event. The
   field of view of the AIA images is $301\arcsec\times229\arcsec$ and the image has been
   rotated. The EUVI data are from the B spacecraft. The dotted line indicates the position of
   the limb as viewed from Earth. The size of the EUVI field of view is
   $409\arcsec\times409\arcsec$. The electronic version of the manuscript contains a movie of
   the running difference images.}
 \label{fig:stereo1}
 \end{figure*}

\begin{figure*}[h!]
 \centerline{\includegraphics[clip,scale=0.45]{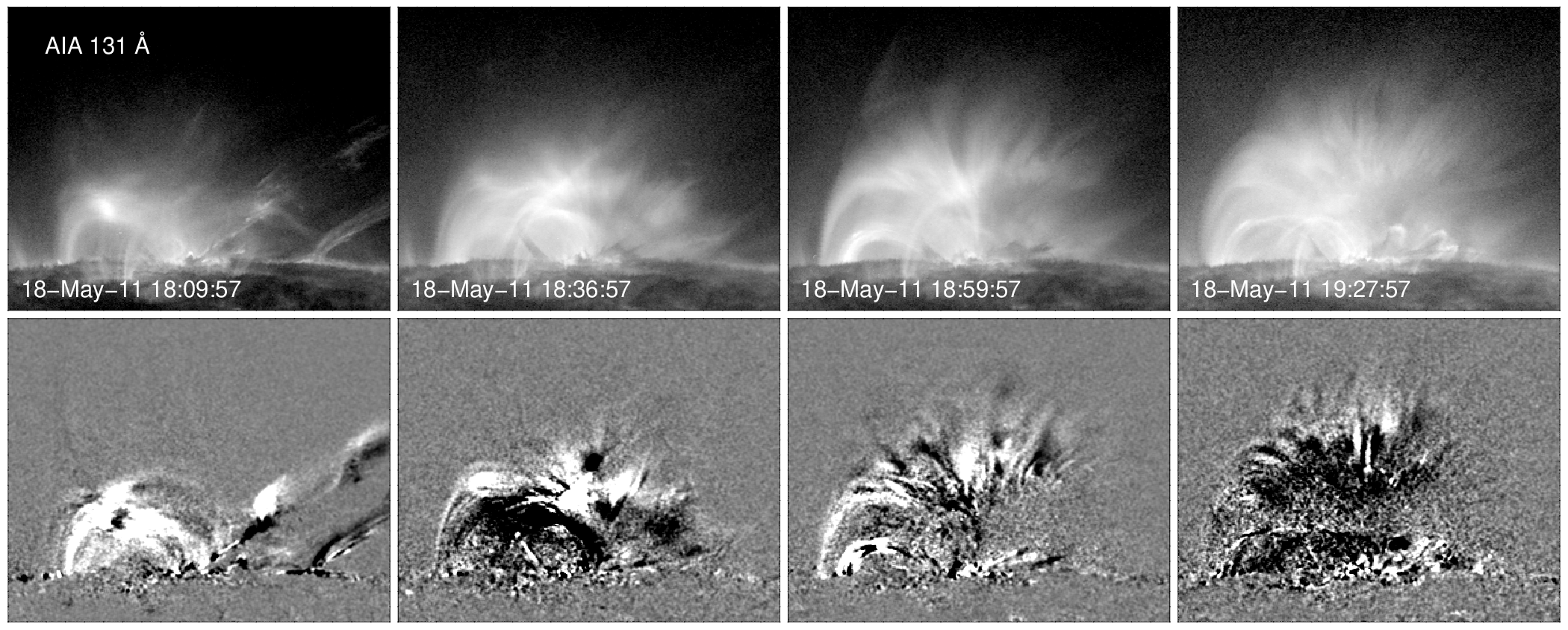}}
 \centerline{\includegraphics[clip,scale=0.45]{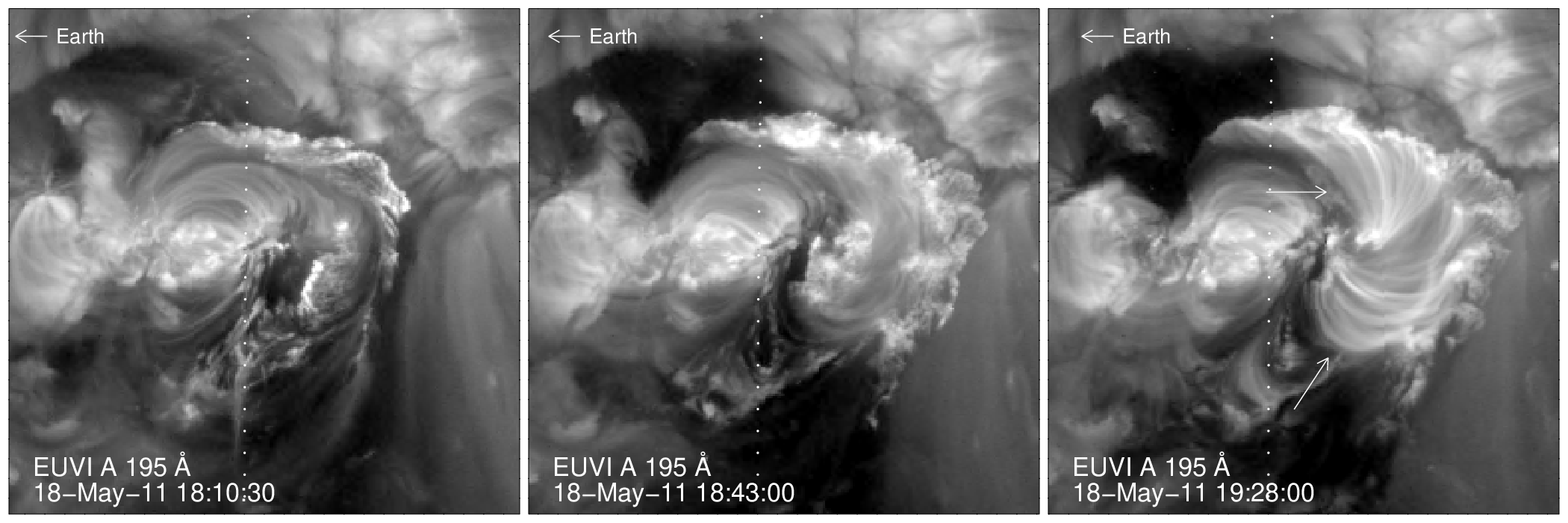}}
 \caption{The same as in Figure \protect{\ref{fig:stereo0}}, but for the 2011 May 18 event. The
   field of view of the AIA images is $277\arcsec\times187\arcsec$ and the image has been
   rotated. The EUVI data are from the A spacecraft. The dotted line indicates the position of
   the limb as viewed from Earth.The size of the EUVI field of view is
   $408\arcsec\times408\arcsec$. The electronic version of the manuscript contains a movie of
   the running difference images.}
 \label{fig:stereo2}
 \end{figure*}

 Some representative AIA 131\,\AA\ images are shown in Figures~\ref{fig:stereo0},
 \ref{fig:stereo1}, and \ref{fig:stereo2}. All three events are clearly associated with a
 coronal mass ejection and have the classic morphology of a two-ribbon flare. Each event shows
 a diffuse ``cloud'' of high temperature ($\sim$10\,MK) plasma that rises with time. Over time
 relatively narrow, lower temperature ($\sim$1\,MK) loops form at the lowest heights and
 material is observed to drain back down to the surface of the Sun along these loops. The
 temperatures that we give here are approximate and based on comparisons with previous work
 \citep[e.g.,][]{warren1999}. High temperature emission is observed only in the 94 and
 131\,\AA\ channels while only low temperature loops are observed in channels such as 171\,\AA,
 which has a significant response only at low temperatures. Very little \ion{Fe}{24} emission
 is observed in the 193\,\AA\ channel images for these flares. The \textit{GOES} soft X-ray
 fluxes for these events are approximately 100 times smaller than was observed in the X-class
 flares considered previously \cite[e.g.,][]{sheeley2004,asai2004}, so this is not surprising.

 \begin{figure*}[h!]
 \centerline{\includegraphics[clip,scale=0.4]{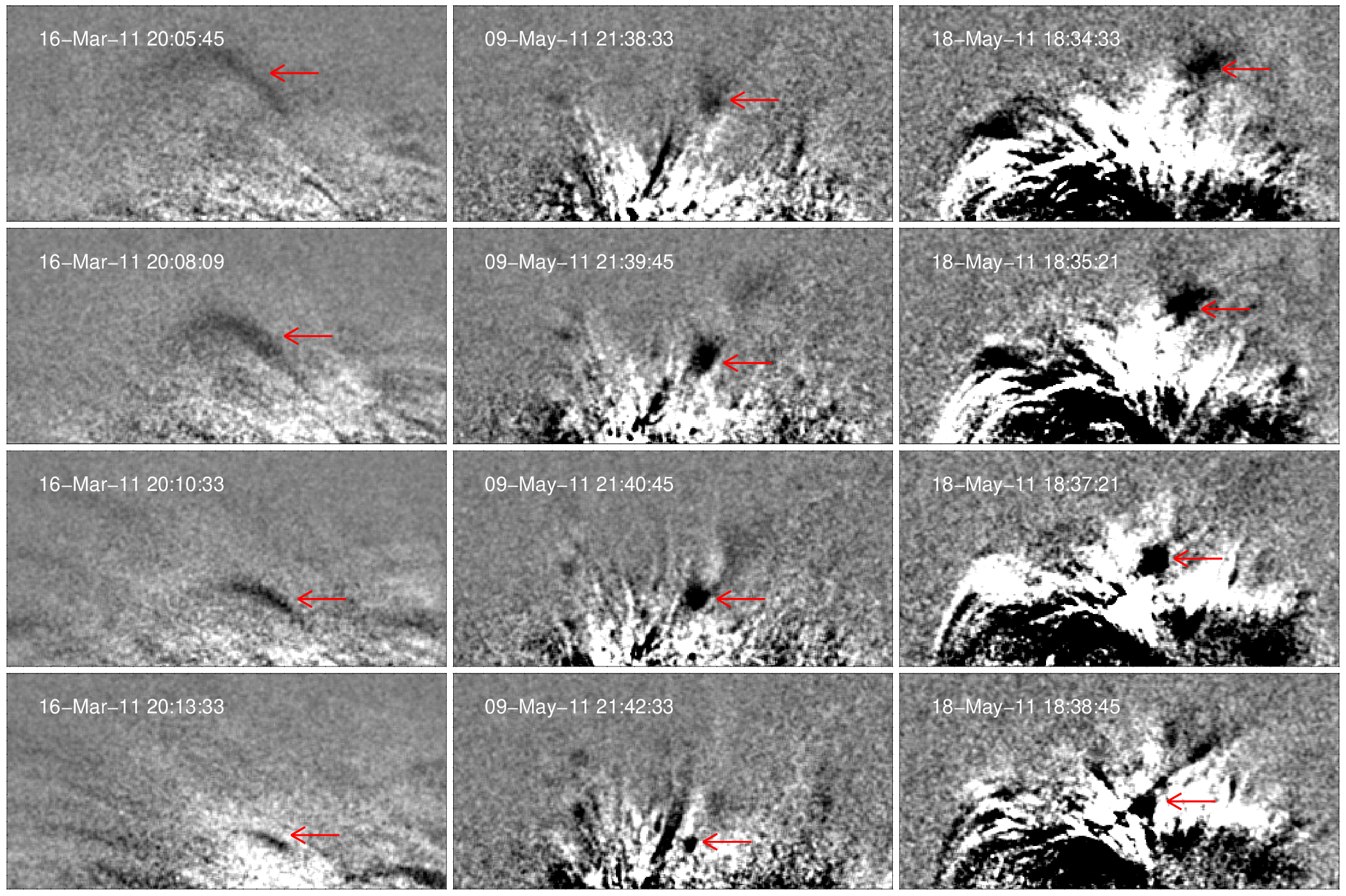}}
 \caption{Example inflows observed with AIA. The left panels shows a descending loop-like
   feature from the March 16 event. The middle panels shows a descending void from the May 9
   flare. The right panels are from the May 18 event.}
 \label{fig:examples}
 \end{figure*}

 Movies of the data for each event show evidence for features forming at large heights and
 descending onto the accumulating post-flare loop arcade. These features are generally
 difficult to track in the raw intensity images. To eliminate the contribution of the more
 slowly varying emission it is useful to compute running difference images.  With the high
 cadence of AIA we have considerable flexibility in constructing these images. Through trial
 and error we have found that averaging sets of 7 images taken 72\,s apart produced the best
 results. Since the downflows move at relatively high velocities, too much averaging or too
 long an interval between frames blurs the features. Averaging fewer frames or choosing a
 shorter interval between frames leads to noisier difference images. Example running difference
 images are displayed in Figures~\ref{fig:stereo0}, \ref{fig:stereo1}, and
 \ref{fig:stereo2}. Movies of the running difference images, which are available in the
 electronic version of the manuscript, show the downflows much more clearly than the raw
 intensity images. In general, the downflows appear as dark features in the running difference
 images, indicating that they are dark relative to the emission into which they are
 descending. 

 \subsection{Morphology of the Downflows}

 In the March 16 event all of the downflows appear as elongated loop-like features. As
 illustrated in the example shown in Figure~\ref{fig:examples}, when the loops first appear
 they are relatively broad but they are compressed as they descend. Many of the descending
 loops that are the easiest to observe appear relatively late in the event (after approximately
 20:00 UT). At earlier times, 18:20--19:00 UT for example, high temperature emission appears at
 relatively low heights very suddenly and does not appear to descend from larger heights. The
 loops that appear near 18:30 UT and are shown in the first column of Figure~\ref{fig:stereo0}
 illustrate this. There is, however, some downward motion after the emission appears,
 consistent with the shrinkage of post-reconnection loops \cite[e.g.,][]{forbes1996}.

 Some of the downflows in these events are observed as dark voids during the entirety of their
 descent. This is true for almost all of the downflows observed in the May 9 event. As with the
 loops observed in the March 16 event, the voids are relatively wide when they appear and are
 compressed in size as they descend. An example downflow is shown in
 Figure~\ref{fig:examples}. The timing of the downflows is similar to that of the March 16
 event, with many of the most easily observed downflows being observed somewhat later in the
 evolution of the flare. Many downflows are observed after 21:15 UT, which is after the peak of
 the \textit{GOES} soft X-ray flux.

 In the May 18 event a mixture of descending voids and loops is observed. For this flare there
 are also some downflows that evolve from voids into loops. An example of a downflow that
 begins as a void and evolves into a loop is shown in Figure~\ref{fig:examples}.  The earliest
 high temperature emission in this event is also observed to appear very suddenly at low
 heights without any downward motion as the loops brighten. After these loops appear, they do
 to contract downward to lower heights. As in the March 16 flare, most of the downflows that
 are easily observed appear after the peak of the soft X-ray emission.

 The \textit{STEREO} observations suggest that the orientation of the May 18 event is
 particularly interesting. The third panel of Figure~\ref{fig:stereo2} shows that the
 post-flare loop arcade forms an ``$\Gamma$'' shape, with part of the ``$\Gamma$'' oriented
 parallel to the line of sight and another part oriented perpendicular to it. When the axis of
 the arcade is along the line of sight the downflows appear as loops or as voids that
 transition to loops. \textit{STEREO} observations indicate that the other events are
 consistent with this interpretation. The arcade in the March 16 event is oriented at an angle
 to the line of sight and loops are observed (see Figure~\ref{fig:stereo0}). The arcade in the
 May 9 event is essentially perpendicular to the line of sight and only voids are observed (see
 Figure~\ref{fig:stereo1}).

 For each event we also inspected the 94\,\AA\ images. As noted earlier, this channel includes
 a strong \ion{Fe}{18} emission line and is sensitive to somewhat lower temperature plasma than
 the 131\,\AA\ channel. As one would expect, these images tend to show the somewhat cooler
 plasma at the lower part of the arcade and as a result tend to show the terminal phase of the
 downflows. For this reason we focus on the data from the AIA 131\,\AA\ channel in this paper.

 \subsection{Height-Time Maps}
 
\begin{deluxetable}{rrcrr}
\tablewidth{2.75in}
\tabletypesize{\small}
\tablecaption{Inflow Properties}
\tablehead{
   \multicolumn{1}{c}{Track}  &
   \multicolumn{1}{c}{$h_0$}  &
   \multicolumn{1}{c}{$a_0$}  &
   \multicolumn{1}{c}{$\tau$} &
   \multicolumn{1}{c}{$v_0$}  \\
   \multicolumn{1}{c}{}  &
   \multicolumn{1}{c}{(Mm)}  &
   \multicolumn{1}{c}{(km s$^{-2}$)}  &
   \multicolumn{1}{c}{(s)} &
   \multicolumn{1}{c}{(km s$^{-1}$)}  
}
\startdata
\multicolumn{5}{l}{16 March 2011} \\
          00 &       132.6 &       0.039 &       831.7 &       -32.1 \\
          01 &       149.0 &       2.282 &       119.0 &      -271.6 \\
          02 &       178.6 &       0.274 &       454.1 &      -124.5 \\
          03 &       196.6 &       0.492 &       290.6 &      -143.0 \\
          04 &       204.3 &       0.857 &       226.7 &      -194.2 \\
          05 &       136.5 &       0.619 &       261.2 &      -161.7 \\
          06 &       188.9 &       0.271 &       500.6 &      -135.8 \\
          07 &       180.9 &       0.308 &       507.7 &      -156.2 \\
          08 &       225.6 &       0.089 &      1031.3 &       -91.7 \\
          09 &       250.4 &       0.044 &      1938.6 &       -84.6 \\
\multicolumn{5}{l}{9 May 2011} \\
          00 &       113.6 &       2.185 &       128.1 &      -279.9 \\
          01 &       118.9 &       1.043 &       212.5 &      -221.7 \\
          02 &       131.1 &       1.506 &       166.6 &      -250.9 \\
          03 &       117.6 &       0.581 &       231.3 &      -134.3 \\
          04 &       128.2 &       0.440 &       345.3 &      -152.1 \\
          05 &       110.8 &       0.161 &       406.1 &       -65.3 \\
          06 &        63.8 &       0.087 &       659.1 &       -57.1 \\
          07 &        82.0 &       0.237 &       363.8 &       -86.1 \\
          08 &        80.4 &       1.050 &       115.6 &      -121.3 \\
          09 &       103.7 &       0.786 &       192.2 &      -151.1 \\
          10 &        73.8 &       0.306 &       245.1 &       -75.1 \\
          11 &        90.9 &       0.375 &       270.7 &      -101.7 \\
\multicolumn{5}{l}{18 May 2011} \\
          00 &        75.2 &       2.687 &       102.1 &      -274.4 \\
          01 &        84.7 &       2.726 &        97.1 &      -264.8 \\
          02 &        83.3 &       0.451 &       274.1 &      -123.7 \\
          03 &        70.5 &       3.407 &        77.7 &      -264.8 \\
          04 &        61.7 &       1.190 &       120.0 &      -142.8 \\
          05 &        75.1 &       0.867 &       165.8 &      -143.7 \\
          06 &        88.3 &       0.473 &       316.6 &      -149.9 \\
          07 &        70.7 &       1.059 &       155.9 &      -165.0 \\
          08 &        76.7 &       0.438 &       243.4 &      -106.5 \\
          09 &        83.1 &       0.403 &       440.2 &      -177.0 
\enddata
\label{table:inflows}
\end{deluxetable}

 \begin{figure*}
 \centerline{\includegraphics[clip,scale=0.45]{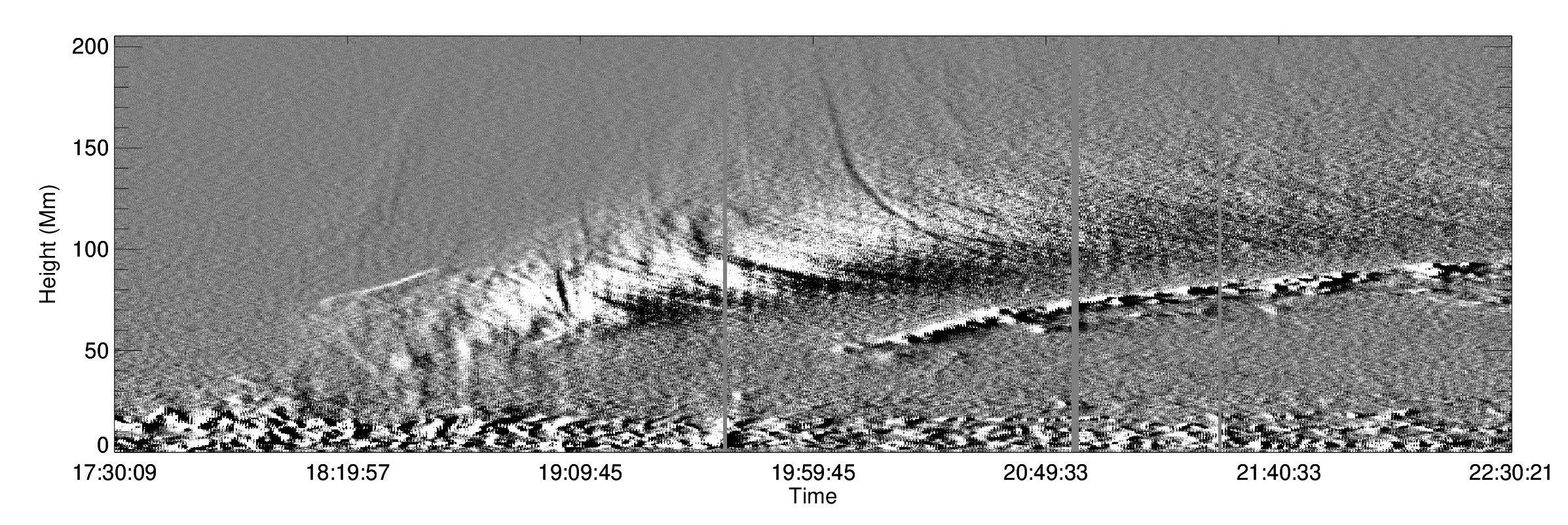}}
 \centerline{\includegraphics[clip,scale=0.70]{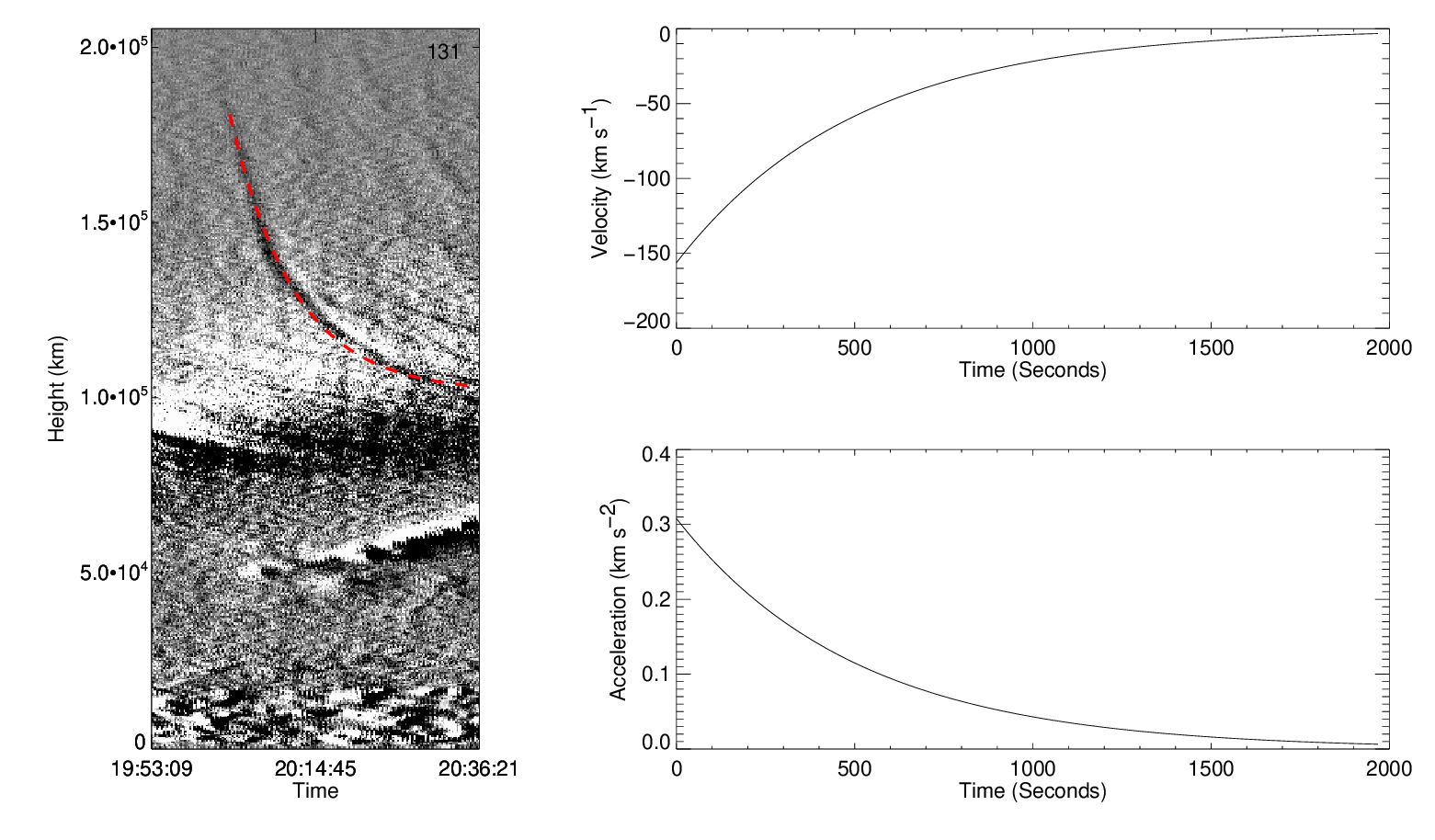}}
 \caption{An example AIA 131\,\AA\ height-time map.  The top panel shows the height-time map
   for most of the event. This figure illustrates some basic features common to all of the
   events: The rapid expansion of the hot plasma (about 40\,km~s$^{-1}$), the dark tracks of
   the inflows that descend into the hot plasma, and the formation of cool, post-flare loops at
   the bottom of the arcade. The rise of the cool post-flare loops is initially rapid
   (10\,km~s$^{-1}$), but gradually decelerates to about 2\,km~s$^{-1}$ toward the end of the
   observations. The bottom panels show an expanded view of the downflow of interest and the
   kinematic information derived from the height-time track. This is track number 7 from 2011
   March 16 (see Table~\protect{\ref{table:inflows}}).}
 \label{fig:ht0}
 \end{figure*}

 A useful way of tracking features as they descend is to create height-time maps. These maps
 are formed by extracting a narrow ``slit'' from each image and stacking them horizontally as a
 function of time \citep[see][]{sheeley2004}. Interpolation is used to extract intensities
 at arbitrary angles. An example height-time map computed from the difference images is shown
 in Figure~\ref{fig:ht0}. Here the downflows are evident as dark tracks, indicating that the
 downflows are generally less intense than the emission into which they are descending. 

 To characterize the kinematics of the downflows we fit selected height-time tracks to a
 function of the form
 \begin{equation}
   h(t)  =  h_0 + v_Tt + a_0\tau^2(e^{-t/\tau}-1).
   \label{eq:h}
 \end{equation}
 With this assumption the velocity of the downflow as a function of time is
 \begin{equation}
   v(t)  =  v_T - a_0\tau e^{-t/\tau},
 \end{equation}
 and the acceleration is
 \begin{equation}
   a(t)  =  a_0 e^{-t/\tau}.
 \end{equation}
 This functional form for the height captures the rapid transition from a relatively high
 initial velocity of $v(t=0)\approx -a_0\tau$ at a height of $h_0$, to a relatively slow
 terminal descent, $v_T$. Note that for features that descend at progressively slower speeds
 the initial velocity will be negative and the acceleration will be positive. 

 We have selected approximately 10 downflows from each event and traced out their tracks in the
 height-time maps. The best-fit parameters for Equation~\ref{eq:h} were determined for each
 track using a least-squares algorithm and are given in Table~\ref{table:inflows}. Each track
 is also indicated in the movies included in the electronic version of the manuscript. 

 We tended to select high velocity tracks that appeared at relatively large heights since they
 were easier to track and are more relevant to the initial dynamics of post-reconnection flux
 tubes. Many of the tracks that we selected decelerate quickly and could not be followed for
 long periods of time. This means the asymptotic behavior is not well observed for these
 tracks. If we leave $v_T$ as a free parameter in the fitting of these tracks the terminal
 velocity often converges to relatively large number, inconsistent with the general trend in
 the data. For tracks that do have a well observed terminal phase, the terminal velocity is
 typically about 4--5\,km~s$^{-1}$. Since the terminal velocities are generally small and we are
 more interested in the initial dynamics we have forced $v_T=0$ for the final fits given in
 Table~\ref{table:inflows}.

 With the improved imaging capabilities of AIA we anticipated being able to track downflows at
 larger heights above the solar surface. Comparing the values for $h_0$ given in
 Table~\ref{table:inflows} with Figures 6 from \cite{savage2011} we see that is the case. Our
 median initial height is approximately 111\,Mm compared with 82\,Mm with the earlier data. Our
 initial velocities and accelerations, however, are very similar to those presented by
 \cite{savage2011} in their Figure 5. For the initial velocity we have a median value of
 -144\,km~s$^{-1}$ compared with -146\,km~s$^{-1}$ in \cite{savage2011}. Our median initial
 acceleration is 0.68\,km~s$^{-2}$ compared with 0.42\,km~s$^{-2}$. Thus it appears that small
 initial velocities relative to the Alfv{\'e}n speed are not an artifact of the lower heights
 at which the downflows have been observed previously.

 Chromospheric evaporation is an essential component of solar flares models based on magnetic
 reconnection \cite[e.g.,][]{fisher1985}. The release of energy on reconnected field lines will
 ultimately lead to the heating and evaporation of chromospheric material. This should lead to
 loops that brighten over time and numerical simulations suggest that this process should occur
 rapidly \cite[e.g.,][]{mariska1989}. One limitation of the running difference images is that
 they can obscure the evolution of the absolute intensities. The downflows are dark features in
 the running difference images, indicating that they have lower intensities than the
 surrounding emission. It is not clear, however, if they remain dark in an absolute sense or
 brighten over time.

 To address this issue we have followed the intensities along each track in the unsubtracted
 images. For comparison we also determined intensities at several locations outside of the
 track. In all cases the intensity in the center of the feature simply tracks the background
 intensities measured outside of the downflow. There is no evidence for the intensities in the
 downflows rising faster than the background. Unfortunately, these measurements are difficult
 to make and subject to considerable uncertainty. The downflows have a very low contrast
 relative to the ambient emission in the flare. The difference images shown in
 Figures~\ref{fig:stereo0}--\ref{fig:ht0} are all scaled over a range of about
 $\pm$5\,DN~s$^{-1}$, which is typically only a few percent of the absolute intensity.


 \section{Discussion}

 We have presented the first observations of supra-arcade downflows with the AIA instrument on
 \textit{SDO}. With the new capabilities of AIA we are able to track inflows at much larger
 heights and for much smaller events than in previous observations. These observations provide
 many new examples of features forming at large heights and descending onto the accumulating
 post-flare loop arcade. The dynamics and morphology of the downflows provides compelling
 evidence for the role of magnetic reconnection in eruptive solar flares. 

 Despite the qualitative agreement between the observations and the standard picture of
 magnetic reconnection, there are several problems that remain. The low velocities that are
 observed for these outflows, for example, are incompatible with the simple 2D models of
 Sweet-Parker and Petschek reconnection, which have Alfv{\'e}nic velocities for the
 outflows. \cite{linton2006} suggest that the interaction of the reconnected flux tubes with
 pre-existing magnetic fields provide a drag force that slows the downflows.

 Another difficulty in interpreting the downflows as evidence for magnetic reconnection is the
 formation of the initial flare plasma. For most events the downflows that are easily observed
 occur late in the flare and descend into an existing cloud of hot emission. The initial flare
 emission forms essentially ``in place,'' and not as a descending loop --- although there is
 some evidence for collapsing motions after these loops appear.

 Finally, the reconnection of magnetic fields should lead to the release of energy that drives
 chromospheric evaporation.  All of the downflows that we detect here, however, are observed as
 depletions relative to the ambient high temperature emission into which they are
 descending. Recent work by \cite{guidoni2011} suggests that density depletions can be formed
 in reconnecting flux tubes that move through a nonuniform current sheet. Their model, however,
 does not include chromospheric evaporation and it is not clear how long such density
 depletions would persist.
 
 The inability of the simple, two-dimensional reconnection models to describe many details of
 the downflows suggests that three-dimensional modeling is required. The AIA observations we
 have shown here give ample evidence for inhomogeneity in the reconnecting magnetic fields and
 this inhomogeneity will be a component of any successful model. High resolution observations
 of the photospheric magnetic field in active regions, such as those taken with the Solar
 Optical Telescope (SOT) on \textit{Hinode} \cite[e.g,][]{kubo2007}, show that the magnetic
 field is structured on the spatial scale of the granules ($\sim$1000\,km) and we conjecture
 that this inhomogeneity is reflected in the highly structured appearance of the flare plasma.


 \acknowledgments This work was supported by NASA and the Office of Naval Research. The authors
 would like to thank Silvina Guidoni for interesting discussions on magnetic reconnection in
 three dimensions.


\end{document}